\documentclass{PoS}
	\usepackage{feyn}
	\usepackage{amsmath}
	\usepackage{braket}
	\usepackage{amsfonts}
	\usepackage{color}
	\usepackage{array}
	\usepackage{multirow}
	\usepackage{graphicx}
	\usepackage{wrapfig}
	\usepackage{caption}
	\usepackage{subcaption}
	\usepackage{tabularx}
	\usepackage{subfig}
	\usepackage{placeins}
	\usepackage{cite}
\usepackage{bm}

\title{Utilising optimised operators and distillation to extract scattering phase shifts}

\ShortTitle{Utilising optimised operators and distillation to extract scattering phase shifts}

\author{\speaker{Antoni J. Woss}\\
       Department of Applied Mathematics and Theoretical Physics, Centre for Mathematical Sciences,
       University of Cambridge, Wilberforce Road, Cambridge, CB3 0WA, UK\\
       E-mail: \email{A.J.Woss@damtp.cam.ac.uk}}

\author{Christopher E. Thomas\\
        Department of Applied Mathematics and Theoretical Physics, Centre for Mathematical Sciences,
        University of Cambridge, Wilberforce Road, Cambridge, CB3 0WA, UK\\
        E-mail: \email{C.E.Thomas@damtp.cam.ac.uk}}

\abstract{(for the Hadron Spectrum Collaboration) \\ \\ In this investigation, we examine how the precision of energy spectra and scattering phase shifts, extracted in lattice QCD, depend upon the degree of distillation type smearing \cite{Peardon:2009gh}. We use the variational method to extract energy spectra for the isospin-1, $J^{PC}=1^{--}$ channel and use the L\"{u}scher method to compute scattering amplitudes, relevant for the $\rho$ resonance, in $\pi\pi$ elastic scattering. Optimised interpolating operators for a single ground state pion are constructed and these are used to construct two pion operators. Calculations are performed on an anisotropic lattice with a pion mass of $m_\pi=236$MeV. We provide a comprehensive comparison of energy spectra and scattering phase shifts across distillation spaces of varying rank.}

\FullConference{34th annual International Symposium on Lattice Field Theory\\
	 24-30 July 2016\\
	 University of Southampton, UK}

\begin{document}

\section{Introduction}
The method for calculating infinite volume scattering amplitudes from finite volume energy spectra has been known since L\"{u}scher pioneered the formalism in the mid 80's \cite{Luscher:1985dn,Luscher:1986pf} but it has not been feasible to apply in practice until recently. To calculate scattering amplitudes precisely, one must extract energy spectra reliably. Above or close to threshold, we must include relevant multi-hadron operators, as demonstrated in figure 6 of \cite{Wilson:2015dqa}. In general, such operators will give rise to Wick contractions with quark annihilations, thus requiring all-to-all propagators in the evaluation of the correlation function. Computing these all-to-all propagators is prohibitively expensive and we use the method of distillation to tackle this issue \cite{Peardon:2009gh}. This method projects quark fields into a subspace, ``distillation space'', of low energy modes and for a sufficiently low rank subspace, computation of all-to-all propagators in this space is made possible.

One motivation behind this study is to perform precision calculations for small pion masses. Lattice calculations of energy spectra and scattering phase shifts towards the physical pion mass have historically been a challenge. Undesirable finite volume effects become large, for example $\exp (-m_\pi L_s)$ where $m_\pi$ is the pion mass and $L_s$ is the lattice spatial extent. By moving to larger spatial volumes, these effects can be put back at the sub-percent level. 
However, the rank of the distillation space needed to maintain a comparable low energy projection scales approximately linearly with spatial volume \cite{Peardon:2009gh} and the computation including all-to-all propagators once again becomes prohibitively expensive. In this study, we look in more detail at how precision calculations depend on the size of the distillation space.

The $\rho$ resonance is reliably extracted and well understood on the lattice and makes for a good testing ground \cite{Wilson:2015dqa}. The following investigation compares energy spectra and scattering phase shifts extracted for $\pi\pi$ elastic scattering on a single volume with different rank distillation spaces. We demonstrate, depending on the quantities to be computed, the capability a small rank distillation space has on producing precision results.

\section{Background} \label{background}
\subsection{Distillation} \label{D}
The method of distillation \cite{Peardon:2009gh} enables calculations with quark annihilations at both source and sink by vastly reducing the size of the propagators. This is done through a projection of the quark fields into a low energy subspace, distillation space. In distillation space, if the rank is small enough, all the required contributions to the correlation function, including quark annihilations, are computationally affordable and can be explicitly evaluated. Within this subspace, one can also affordably project operators onto some definite momentum at source and sink. 

Consider the Dirac propagator on a single time slice and single spinor degree of freedom. It has dimension $M=(L_s/a_s)^3\times N_c$, where $a_s$ is the spatial lattice spacing and $N_c$ is the number of colours. For a large spatial volume, e.g. $32^3$ in this study, this $M \times M$ matrix is large, $M\sim10^5$, and it is unfeasible to include all-to-all propagators.   

Distillation is the procedure of projecting onto the lowest $N$ eigenvectors $\xi_x^{(k)}(t)$, ordered by magnitude of eigenvalue, of the gauge covariant lattice Laplacian, $-\nabla^2_{xy}(t)$. This has the effect of smearing out high energy quark and gauge field fluctuations. We write ``distillation space of rank $N$'' to refer to the number of eigenvectors used in the projection.
We project the quark fields, $\psi_x \rightarrow \square_{xy} \psi_y$, where the projection operator, $\square_{xy}(t)$, at some fixed time $t$ is\footnote{Subscripts $x,y$ run over colour and spatial sites, i.e. from $1$ to $M$.},
\begin{equation} \label{eq:1}
\square(t) = V(t)V^\dagger(t) \Longrightarrow \square_{xy}(t)=\sum_{k=1}^{N}\xi_x^{(k)}(t) \xi_y^{(k)\dagger}(t)
\end{equation}
Previous calculations have set a bench mark for the number of vectors one should use on a given volume; figure 9 reference \cite{Dudek:2010wm}. Crucially, in all instances, rank $N \ll M$, and for example in this study $32 \leq N \leq 384$. Projected propagators and operators\cite{Peardon:2009gh} can be stored cheaply and for mesons evaluating the correlation functions reduces to matrix products and traces in an $N\times N$ vector space. 
\subsection{Variational Method}\label{VM}
Fermion bilinear operators that resemble a meson, $\bar{q}\{\Gamma \overleftrightarrow{D} ... \overleftrightarrow{D}\} q$, can be constructed to transform within some lattice irreducible representation, ``irrep'', by appropriately coupling products of Dirac $\gamma$ matrices, $\Gamma$, with covariant derivatives, $\overleftrightarrow{D}$\cite{Dudek:2010wm}. We refer to these as $\bar{q}q$ type operators and can project these onto some overall momentum \cite{Thomas:2011rh}. For a basis of operators, $\{O(t)_i\}$, transforming within some lattice irrep, we construct the correlation matrix $C_{ij}(t,0)=\braket{O(t)_iO^\dagger(0)_j}$. The variational method \cite{Luscher:1990ck,Michael:1985ne} allows the extraction of the energies of the states in the theory. This is done by solving the generalized eigenvalue problem,
\begin{equation} \label{eq:6}
C(t)v^n(t)=\lambda_n(t,t_0)C(t_0)v^n(t)
\end{equation}
where $\lambda_n(t_0,t_0)=1$ and by construction $v^{m\dagger}(t)C(t_0)v^n(t)=\delta_{mn}$. $t_0$ is chosen appropriately, as demonstrated in \cite{Dudek:2007wv}.
It can be shown that at large times the principal correlators, $\lambda_n(t,t_0)$, behave as $\lambda_n(t,t_0)=e^{-E_n(t-t_0)}(1+\text{O}(e^{-|\delta E|(t-t_0)}))$ where $E_n$ is the energy of state $n$ and $\delta E$ is the energy gap to the nearest state to $n$. In our implementation, principal correlators are fitted to two exponentials, the fit function being $\lambda_n(t,t_0)=(1-A_n)e^{-E_n(t-t_0)}+A_ne^{-E'_n(t-t_0)}$, with $E_n$, $E'_n$ and $A_n$ fit parameters. The second exponential accounts for the next to leading order term. For energies we extract, in any irrep, to be a reliable determination of the actual energy eigenvalues, we must include enough operators with different structures so as to disentangle the low lying eigenstates. 

The eigenvector, $v^n(t_Z)$, gives a set of coefficients for the \textit{optimal operator}\footnote{This $t_Z > t_0$ is some reference time slice on which we take coefficients \cite{Dudek:2007wv}.}, the optimal linear combination of operators from  $\{O(t)_i\}$, $O(t)_{opt,n}=\sum_{i}v^n(t_Z)_iO(t)_i$, that interpolates the $n^{th}$ state \cite{Dudek:2012gj}.

For scattering processes, it is essential to not only include operators that are  $\bar{q}q$ type but also operators resembling meson-meson when such meson-meson thresholds are close by or kinematically open. 

In this study we include meson-meson $\pi\pi$ operators as well as $\bar{q}q$ operators, both with overall momentum $\vec{P}$. The $\pi\pi$ operators are built from the product of two $\pi$ operators, $\chi(\vec{p},\Lambda)$,

\begin{equation}
\pi\pi(\vec{P},\Lambda)=\sum_{\vec{p_1}+\vec{p_2}=\vec{P}}C(\Lambda,\Lambda_1,\Lambda_2;\vec{P},\vec{p_1},\vec{p_2})\chi(\vec{p_1},\Lambda_1)\chi(\vec{p_2},\Lambda_2)
\end{equation}
Here $\chi(\vec{p},\Lambda)$ is the optimal operator for interpolating a ground state pion. The sum is over momenta related by lattice rotations and is weighted by lattice Clebsch Gordan coefficients, $C$, that project the two-meson state onto the correct irreducible representation $\Lambda$ \cite{Dudek:2012gj}. 
\subsection{L\"{u}scher's method} \label{LM}
L\"{u}scher's method \cite{Luscher:1985dn,Luscher:1986pf,Luscher:1990ux,Christ:2005gi,Kim:2005gf,Luscher:1991cf} relates discrete finite volume energy levels to infinite volume scattering amplitudes through a quantization condition. We can summarise the elastic scattering quantization condition with the determinant equation,
\begin{equation} \label{detcond}
\det [\bm{1}+i\bm{\rho}(E_{cm}) \cdot \bm{t}(E_{cm}) \cdot (\bm{1}+i\bm{M}(q^2))] = 0
\end{equation} 
where the matrices are in the space of partial waves $l$. $\bm{t}$ is the infinite volume scattering \textit{t}-matrix, related to the \textit{S}-matrix via $\bm{S}=\bm{1}+2i\sqrt{\bm{\rho}}\cdot\bm{t}\cdot \sqrt{\bm{\rho}}$, and is diagonal in partial wave for $\pi\pi$ scattering with components $t^{(l)}$. $\bm{\rho}$ is a diagonal matrix with entries $\rho_{ll'}(E_{cm})=2k\delta_{ll'}/E_{cm}$ where $k=|\vec{k}|$ is the 3-momentum in the cm frame and $E_{cm}$ is the energy in the cm frame. The matrix $\bm{M}(q^2)$, with $q=kL/2\pi$, is off diagonal in $l$ and encodes effects of a finite volume such as partial wave mixing. It is a known function of spherical harmonics and generalised $\zeta$-functions -- see \cite{Rummukainen:1995vs} equation (89).
\section{Results}
\subsection{Spectra}
Calculations were performed on $\sim480$ gauge field configurations of volume $(L/a_s)^3\times(T/a_t)=32^3\times256$, with spatial lattice spacing $a_s\sim0.12\text{ fm}$, temporal lattice spacing $a_t=a_s/\xi$ with $\xi\sim3.5$ and a pion mass, $m_{\pi}=236\text{ MeV}$, see \cite{Wilson:2015dqa} for details. We calculate the correlation function with $\bar{q}q$ and $\pi\pi$ operators transforming in the allowed lattice irreps for isospin-1 and $J^{PC}=1^{--}$. Operators are constructed according to reference \cite{Wilson:2015dqa} with the benefit of optimised operators demonstrated in figure 2 of \cite{Dudek:2012gj}. We average over two time sources and for $N=64$, $96$ analyse only $[000]T_1^-$ and $[\vec{P}]A_1$-type irreps. Table \ref{table3} gives the number of fermion bilinear operators and $\pi\pi$ operators used\footnote{Note that we do not include $K\bar{K}$ and $\pi\pi\pi\pi$ like operators because we restrict our attention to the elastic $\pi\pi$ region.}.

\begin{figure}[h!]
	 \vspace{-0.4cm}
     \centering
     \centering
       \includegraphics[width=1\linewidth]{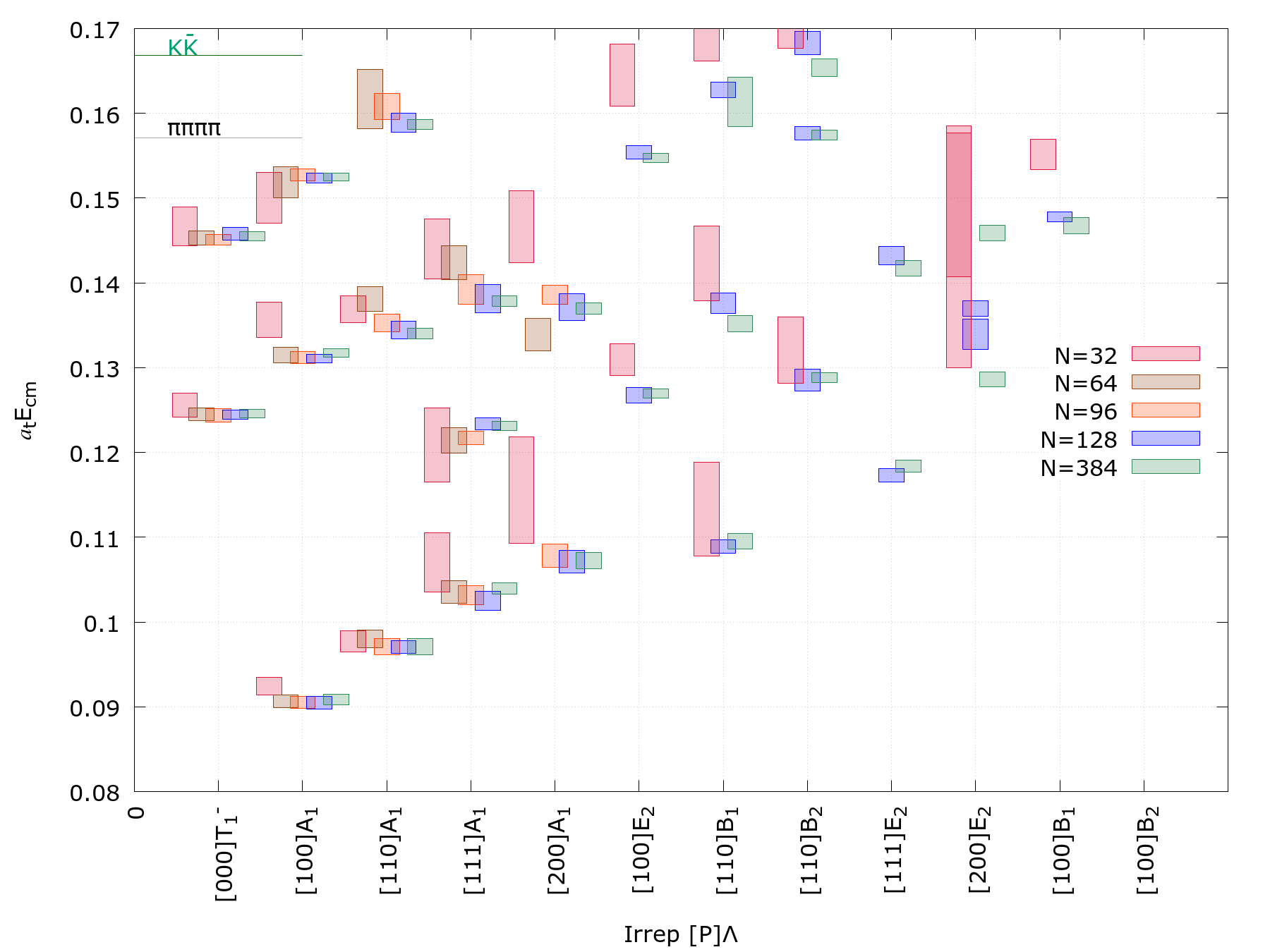}
     \caption[Caption title in LOF]{Energy levels (cm energies in lattice units) for various overall momenta $\vec{P}$ and irreps $\Lambda$, for distillation spaces of various ranks, $N$. The grey horizontal line is the $\pi\pi\pi\pi$ threshold and green line the $K\bar{K}$ threshold. Since we do not include $\pi\pi\pi\pi$ or $K\bar{K}$ like operators, spectra are reliable only below these thresholds.}
     \label{fig:22}
     \vspace{-0.4cm}
     \end{figure}

We compare extracted spectra in figure \ref{fig:22} for the twelve different overall momenta and irreps, $[\vec{P}]\Lambda$, for distillation spaces of rank $N=32,64,96,128$ and $384$. In the final comparisons, we varied the bases of operators to ensure clean signals for the extracted spectra. We note, in particular, the good agreement between all ranks at low momenta and great agreement across the irreps for rank $N=128$ and $384$.
\begin{table}[h!]
\vspace{-0.5cm}
  \small
\begin{center}

 \begin{tabular}{c c c c c c}
 \hline
 $[000]T^-_1$ & $[100]A_1$ & $[110]A_1$ & $[111]A_1$ & $[200]A_1$ & $[100]E_2$ \\ 
 \hline
 $\pi_{001}\pi_{00-1}$ & $\pi_{000}\pi_{100}$ & $\pi_{000}\pi_{110}$ & $\pi_{000}\pi_{111}$ & $\pi_{000}\pi_{200}$ & $\pi_{0-10}\pi_{110}$\\ 
 $\pi_{011}\pi_{0-1-1}$ & $\pi_{-100}\pi_{110}$ & $\pi_{100}\pi_{010}$ & $\pi_{100}\pi_{011}$ & & $\pi_{0-1-1}\pi_{111}$ \\
 $\pi_{111}\pi_{-1-1-1}$ &  & $\pi_{00-1}\pi_{111}$ & $\pi_{-111}\pi_{200}$ & & \\
 &  & $\pi_{101}\pi_{01-1}$ &  &  & \\ 
 \hline
 $\bar{\psi}{\Gamma}\psi \times 26$& $\bar{\psi}{\Gamma}\psi \times 18$ & $\bar{\psi}{\Gamma}\psi \times 27$ & $\bar{\psi}{\Gamma}\psi \times 21$ & $\bar{\psi}{\Gamma}\psi \times 18$ & $\bar{\psi}{\Gamma}\psi \times 29$ \\ 
 \hline
\end{tabular}
 \begin{tabular}{c c c c c c}
 \hline
 $[110]B_1$ & $[110]B_2$ & $[111]E_2$ & $[200]E_2$ & $[100]B_1$ & $[100]B_2$ \\ 
 \hline
 $\pi_{010}\pi_{100}$ & $\pi_{00-1}\pi_{111}$ & $\pi_{100}\pi_{011}$ & $\pi_{1-10}\pi_{110}$ & $\pi_{0-10}\pi_{110}$ & $\pi_{0-1-1}\pi_{111}$  \\
 $\pi_{01-1}\pi_{101}$ & $\pi_{0-10}\pi_{110}$ & $\pi_{-111}\pi_{200}$ & $\pi_{1-1-1}\pi_{111}$ & & \\ 
 \hline
 $\bar{\psi}{\Gamma}\psi \times 29$ &  $\bar{\psi}{\Gamma}\psi \times 29$ &  $\bar{\psi}{\Gamma}\psi \times 35$ &  $\bar{\psi}{\Gamma}\psi \times 29$ &  $\bar{\psi}{\Gamma}\psi \times 9$  &  $\bar{\psi}{\Gamma}\psi \times 9$\\ 
 \hline
\end{tabular}
\centering
\caption{Operators used for each momentum $\vec{P}$ and irrep $\Lambda$, labelled as $[\vec{P}]\Lambda$, in the variational analysis. $\bar{\psi}\Gamma\psi\times N$ denotes the number, $N$, of fermion bilinear operators. $\pi_{\vec{p}_1}\pi_{\vec{p}_2}$ denotes a $\pi\pi$ operator where the $\pi$ momenta $\vec{p_1}$ and $\vec{p_2}$ are summed over as in equation (2.3).}
\label{table3}
\vspace{-0.5cm}
\end{center}
\end{table}
\subsection{Phase Shifts}
\begin{figure}[h!]
     \centering
     \vspace{-0.8cm}
     \captionsetup{justification=,margin=0cm}
     \centering
       \includegraphics[width=1\linewidth]{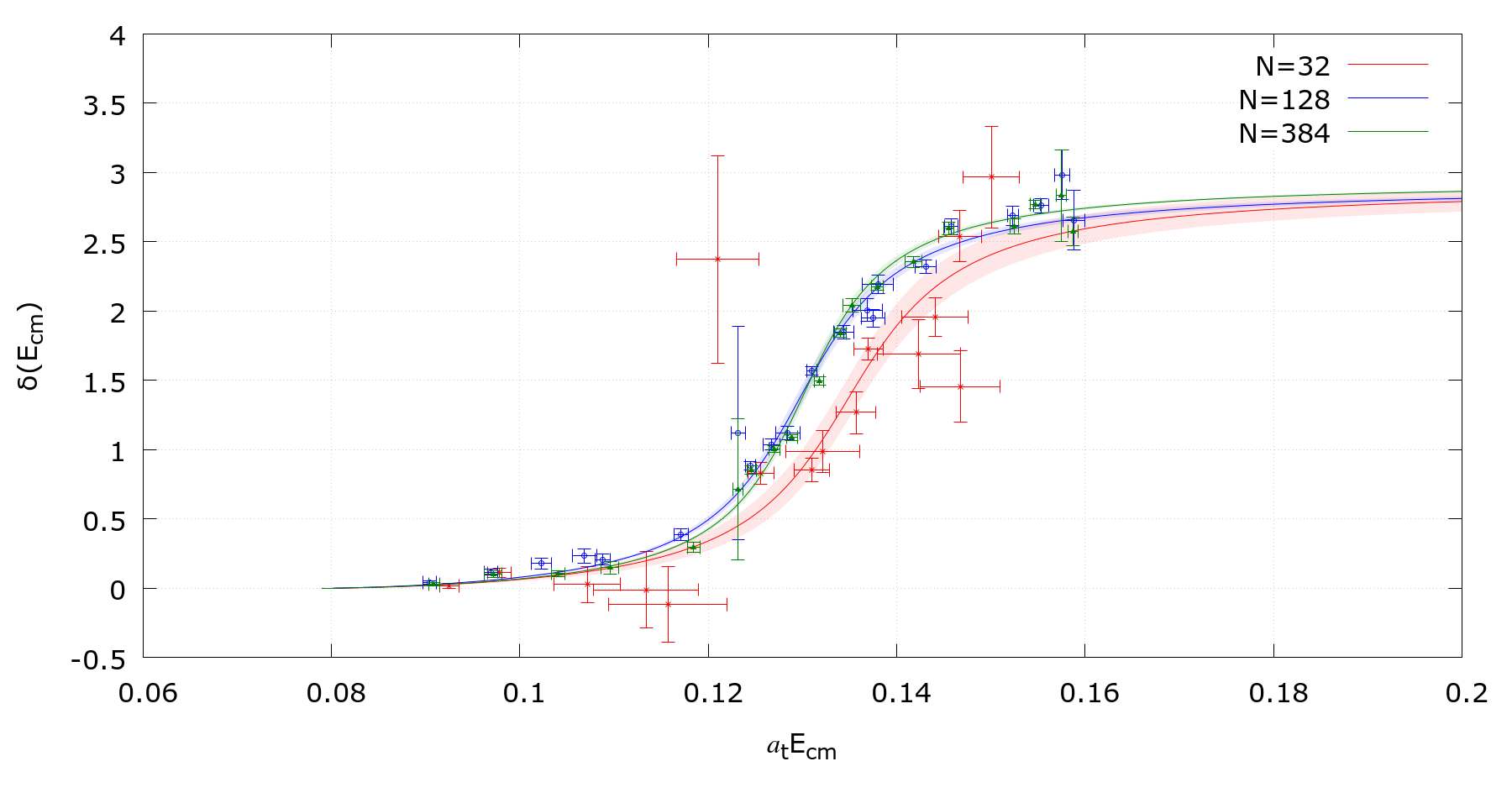}
     \caption{$P$-wave phase shifts as a function of cm energy for various rank distillation spaces. The points show the phase shift at some extracted $E_{cm}$ values. The coloured bands are phase shifts from Breit Wigner parameterisations of $t^{(l)}$ as described in the text (mean and $1\sigma$ uncertainty on either side).}
     \vspace{-0.4cm}
     \label{fig:23}
     \end{figure}

We fit the energy spectrum to the spectrum obtained from a relativistic Breit Wigner parameterisation of $t^{(l)}$, (equation (10) in \cite{Wilson:2014cna}), as described by equation (8) of \cite{Wilson:2014cna}. We consider here only the lowest partial-wave, $P$-wave, as higher partial waves are suppressed, and examine the phase shifts extracted across rank $N=32$, $128$ and $384$ distillation spaces. Figure \ref{fig:23} shows the extracted phase shifts.

We observe in figure \ref{fig:23} very little change in the phase shifts between $N=128$ and $384$ but a noticeable difference when compared with $N=32$. This is what we would anticipate by looking at the differences in the energy spectra, at these ranks, in figure \ref{fig:22}.
\section{Conclusion}
We have demonstrated that the required number of distillation vectors depends on the quantities of interest and for some quantities one can use fewer vectors and still perform precision calculations. This suggests distillation is viable for near or at physical pion mass calculations.

Lattice simulations using a few distillation vectors give an idea of the number of vectors one should use in production runs. We see this for rank $N=32,64$ and $96$ in figure \ref{fig:22}. To obtain a good level of precision for the higher momenta irreps or highly excited states, more vectors are needed, but for the lower momenta irreps, these lower rank spaces are sufficient as demonstrated by the good agreement of such energy spectra across these low rank spaces.
Optimised operators aid in eliminating unwanted excited state contamination at early times and improve the signal to noise ratio for two-meson states. 
\section*{Acknowledgements}
We thank our colleagues within the Hadron Spectrum Collaboration.
AW is supported by the U.K. Science and Technology Facilities Council
(STFC).  CET acknowledges support from STFC [grant ST/L000385/1].
Computations were performed at Jefferson Laboratory under the USQCD
Initiative and the LQCD ARRA project.
The software codes {\tt Chroma}~\cite{Edwards:2004sx}, {\tt
QUDA}~\cite{Clark:2009wm,Babich:2010mu}, {\tt QPhiX}~\cite{Joó2013},
and {\tt QOPQDP}~\cite{Osborn:2010mb,Babich:2010qb} were used to compute
the propagators required for this project.
This research was supported in part under an ALCC award, and used
resources of the Oak Ridge Leadership Computing Facility at the Oak
Ridge National Laboratory, which is supported by the Office of Science
of the U.S. Department of Energy under Contract No. DE-AC05-00OR22725.
This research is also part of the Blue Waters sustained-petascale
computing project, which is supported by the National Science Foundation
(awards OCI-0725070 and ACI-1238993) and the state of Illinois. Blue
Waters is a joint effort of the University of Illinois at
Urbana-Champaign and its National Center for Supercomputing
Applications. This work is also part of the PRAC ``Lattice QCD on Blue
Waters''. This research used resources of the National Energy Research
Scientific Computing Center (NERSC), a DOE Office of Science User
Facility supported by the Office of Science of the U.S. Department of
Energy under Contract No. DEAC02-05CH11231. The authors acknowledge the
Texas Advanced Computing Center (TACC) at The University of Texas at
Austin for providing HPC resources that have contributed to the research
results reported within this paper.
Gauge configurations were generated using resources awarded from the
U.S. Department of Energy INCITE program at the Oak Ridge Leadership
Computing Facility, the NERSC, the NSF Teragrid at the TACC and the
Pittsburgh Supercomputer Center, as well as at Jefferson Lab.

\bibliography{mybib_PoS.bib}
\bibliographystyle{unsrt}

\end{document}